\documentclass[preprint,11pt]{elsarticle}
\usepackage{amssymb}
\usepackage{amsthm,amsmath}
\usepackage{graphicx}
\usepackage{array,multirow}
\usepackage{hyperref}
\usepackage[utf8]{inputenc} 

\journal{arXiv}

\begin{document}

\begin{frontmatter}



\title{An AI-based Solution for Enhancing Delivery of Digital Learning for Future Teachers}

\author[sut2]{Yong-Bin Kang}
\ead {ykang@swin.edu.au}
\author[sut]{Abdur Rahim Mohammad Forkan}
\ead{fforkan@swin.edu.au}
\author[sut]{Prem Prakash Jayaraman}
\ead {pjayaraman@swin.edu.au}
\author[vv]{Natalie Wieland}
\ead {natalie@vidversity.com}
\author[vv]{Elizabeth Kollias}
\ead {liz@vidversity.com}
\author[sut]{Hung Du}
\ead {hungdu@swin.edu.au}
\author[sut]{Steven Thomson}
\ead {101911349@student.swin.edu.au}
\author[mu]{Yuan-Fang Li}
\ead {yuanfang.li@monash.edu}

\address[sut2] {Department of Media and Communication, Swinburne University of Technology, Melbourne, Victoria, Australia}
\address[sut] {School of Science, Computing and Engineering Technologies, Swinburne University of Technology, Melbourne, Victoria, Australia}
\address[vv] {VidVersity, Australia}
\address[mu] {Department of Data Science and AI, Monash University, Melbourne, Victoria, Australia}

\begin{abstract}
There has been a recent and rapid shift to digital learning hastened by the pandemic but also influenced by ubiquitous availability of digital tools and platforms now, making digital learning ever more accessible. An integral and one of the most difficult part of scaling digital learning and teaching is to be able to assess learner’s knowledge and competency. An educator can record a lecture or create digital content that can be delivered to thousands of learners but assessing learners is extremely time consuming.  In the paper, we propose an  Artificial Intelligence (AI)-based solution namely VidVersityQG for generating questions automatically from pre-recorded video lectures. The solution can automatically generate different types of assessment questions (including short answer, multiple choice, true/false and fill in the blank questions) based on contextual and semantic information inferred from the videos. The proposed solution takes a human-centred approach, wherein teachers are  provided the ability to modify/edit any AI generated questions. 
This approach encourages trust and engagement of teachers in the use and implementation of AI in education. The AI-based solution was evaluated for its accuracy in generating questions by 7 experienced teaching professionals and 117 education videos from multiple domains provided to us by our industry partner VidVersity. VidVersityQG solution showed promising results in generating high-quality questions automatically from video thereby significantly reducing the time and effort for educators in manual question generation.
\end{abstract}


\end{frontmatter}

\section{Introduction}
Digital learning has been gaining momentum for the last few decades. However, up until the COVID-19 pandemic caused a seismic shift in the education sector, few educational institutions had fully developed digital learning models in place and adoption of digital models was ad-hoc or only partially integrated alongside traditional teaching modes \cite{laufer2021digital}. 
In the wake of the disruptive impact of the pandemic, the education sector and more importantly educators have had to move rapidly to take up digital solutions to continue delivering learning. At the most rudimentary level, this has meant moving to online teaching through platforms such as Zoom, Google, Teams and Interactive Whiteboards and delivering pre-recorded educational materials via Learning Management Systems (e.g., Echo). Digital learning is now simply part of the education landscape both in the traditional education sector as well as within the context of corporate and workplace learning. 

A key challenge future teachers face when delivering educational content via digital learning is to be able to assess what the learner knows and understands, the depths of that knowledge and understanding and any gaps in that learning. Assessment also occurs in the context of the cohort and relevant band or level of learning. The Teachers Guide to Assessment produced by the Australian Capital Territory Government \cite{act2021teachers} identified that teachers and learning designers were particularly challenged by the assessment process, and that new technologies have the potential to transform existing digital teaching and learning practices through refined information gathering and the ability to enhance the nature of learner feedback.

Artificial Intelligence (AI) is part of the next generation of digital learning, enabling educators to create learning content, stream content to suit individual learner needs and access and in turn respond to data based on learner performance and feedback \cite{cope2020artificial}. AI has the capacity to provide significant benefits to teachers to deliver nuanced and personalised experiences to learners. 
 
One key innovation of AI in the context of assessment process is through Automatic Question Generation (AQG) \cite{das2021automatic}. This is the process of creating questions automatically based on the context of the content being delivered to assess what the learner knows and understands \cite{das2021automatic}.  .  
AQG techniques are useful in reducing the manual efforts devoted by teachers for construction of questions while providing an excellent opportunity to develop an engaging digital learning experience for the learners. However, the use of AQG in digital learning poses several challenges to meet the needs of future teachers. 
First of all, it is necessary to support two categories of questions i.e., \textit{objective} and \textit{subjective}. The objective style asks the learner a question that has a definitive answer. This is traditionally true/false, multiple choice and fill in the blank and is typically the easier style of question to create automatically. The second is the subjective style of question such as short answer or essay style questions. These have traditionally been much harder to simulate through automatic question generators. It is argued that both styles of questions are necessary to provide a complete digital learning experience to learners while enhancing teachers ability to deliver digital learning and related learning outcomes effectively. 
.
In this work, we propose, develop, implement and validate an AI-based solution, VidVersityQG.  VidVersityQG provides future teachers ability for  automatic question generation based on educational video learning content. VidVersityQG takes a human-centric design and development approach that allows the teacher to be in control and where required provide additional inputs into the AI generated questions. The VidVersityQG solution has been jointly developed with our industry partner VidVersity \cite{vidversity2021} who have worked in the development of video based learning creation for a number of years and identified a need to automate the question generation process which is a time consuming component of learning design. This paper explores the role of AI in education, how innovation in AI can deliver a streamlined and nuanced response to learning assessment and an 
evaluation of the outcomes of the use AI together with teacher validation to create an effective and practical digital learning tool. The key contributions of this paper are listed below.

\begin{itemize}
    \item VidVersityQG, a novel AI-based solution for supporting the delivery of digital learning and assessment
    \item Experimental evaluations of VidVersityQG with teachers to validate the efficacy of the solution and the AI-based digital learning and teaching methodology
\end{itemize}

\section{Literature Review}

This section describes the current literature in digital learning and the use of AI for such purpose. It also covers the current state-of-the-art AI techniques for automatic question generation.

\subsection{Digital Learning}
The impact of the pandemic is transformation higher education in terms of widespread, longer term adoption of digital learning as a fully integrated part of the delivery of learning and teaching.

One recent study \cite{jones2020motivating} explored the impact of this rapid transition and the promise of digital learning solutions compared to practical outcomes.  Overall the study found many positive aspects of digital learning and in particular, learners liked the flexibility, interactivity and ability to be self paced. The study in \cite{jones2020motivating} concluded that technologies can positive impact learning outcomes however for the full realisation of the potential benefits principles around learning design are the key drivers and the technology needs to enable this. 

\subsection{AI in Digital Learning}

AI and its application in digital learning has been a particular focus for research in recent years. Initially researchers focused on areas such as chess and speech recognition. AI first started appearing in the education sector in the 1960s with applications such as the natural language processing (NLP) system “Eliza” from MIT \cite{weizenbaum1983eliza}. This first generation of AI really refers to intelligent tutoring systems (ITS) as it was the most common application of AI \cite{ross1987intelligent}. It has been argued that it evoked little interest with educators and they had limited trust around the systems. In the last ten years, research into the use of AI in the education space has developed into a field now as AIEd. AI has in turn evolved from the simple `\textit{tutor}' style model at it’s inception to a highly sophisticated function that is able to synthesise and adapt the most complex of tasks \cite{chatterjee2020adoption}, in particular Deep Neural Network (DNN) based models aka deep learning (DL) \cite{yosinski2014transferable}. 

The widespread adoption of AI within the education sector requires the engagement of all stakeholders from policy makers through to individual teachers. The TAM model (teacher acceptance model) model proposed by \cite{davis1989perceived} was designed to map the effects of external factors on the user belief of how useful the technology will be. This was combined with the internal factors of Self Efficacy and Anxiety, Perceived Usefulness, Perceived Ease of Use and Attitude Towards Use to study how AI has been adopted by teachers \cite{wang2021factors}. The authors in \cite{wang2021factors}  concluded that future studies would need to not only focus on the intersection of the technologies and algorithms with teaching methodologies but also assess any factors that would impact the motivations of teachers to take up these technologies. This goes to the heart of the issue which is that there must be a perceived as well as actual benefit for teachers to adopt AI and utilise the full capability of the technology. It is for this reason that human-centred AI has evolved where the expertise of the teacher combined with the technology can create the most successful AI-based learning and teaching outcomes.

Authors in \cite{yang2021human} recognised that pure technology is different to learning technologies. Learning technologies need to work with humans, they have to interact and teach. They need to have a more human-centred approach. That is, to include the human expert in the process where there is a collaboration between the human and the machine to increase human productivity and output. Not only does this result in a more human centred output it also helps humans (i.e. teachers) adopt and trust new technologies.   
AI, therefore, needs to operate within a human context, to be shaped by human responses and to have an application and benefit that is deeply connected to the human condition and in particular to the nuanced and differing needs of both teachers and students. 

\subsection{AI-based Automatic question generation (AQG)}
The natural question generation is a growing research domain to generate questions from text input. In literature AI, specifically natural language processing (NLP) algorithms are extensively used in AQG \cite{das2021automatic}.  In particular, the advancement of deep learning techniques in NLP \cite{das2021automatic} and readily available natural language dataset (e.g, SquAD \cite{rajpurkar2016squad}) expanded the research potential in the AQG area. In NLP, the question generation method can be divided into 3 categories: syntax-based, semantic-based and template-based \cite{yao2012semantics,fabbri2020template}. Pre-trained deep learning models such as BERT \cite{ch2018automatic}, T5 transformer language model \cite{kumar2021deep}, GPT-2 and GPT-3 language model \cite{oniani2020qualitative} are proven highly useful for question generation from text data based on syntax and semantics in several application domains. Various NLP methods are developed in the literature for natural question generation such as the sequence-to-sequence (Seq2Seq), \cite{su2018follow}, encoder-decoder\cite{du2017learning}, and QG-Net \cite{wang2018qg}. In digital learning domain, Authors in \cite{tsai2021automatic}  proposed an NLP-based method that combined syntax-based and semantic-based methods for AQG. The authors evaluated the proposed method on 41 students and identified improved learning experience for students. For generating question answer pairs from transcribed video data authors in \cite{yang2021just} developed HowToVQA69M, dataset with 69M video-question answer triplets. Such pre-trained datasets are highly valuable to develop future NLP techniques for AQG from transcribed text from video.

Most of these state-of-the-art models for question generation use a single sentence as input to generate questions which is not useful for generating questions of educational content available as pre-recorded videos. ParaQG \cite{kumar2019paraqg}, an advanced NLP model is introduced to tackle such problems and capable of generating questions from paragraphs. ParaQG is based on a combination of NLP techniques such as Seq2Seq model with dynamic dictionaries, the copy mechanism and the global sparse-max attention. In our work, we adopted ParaQG and have significantly expanded on it capability for AQG. Specifically, the input data for our AI model incorporated in VidVersityQG (described in detailed in next section) is generated from the transcribed text extracted from educational video learning content. Furthermore, our proposed model has the capability to generate both subjective and objective questions.
 
The state-of-the-art techniques utilising AI for AQG have demonstrated significant benefit in automated generation of questions. However, they lack a human-centred approach, and have not been co-designed and/or assessed by teachers/educators who work in the digital learning space. Moreover, most of the developed techniques are tested only on publicly available natural language datasets \cite{tsai2021automatic} and most of these detests are not necessarily suitable for learning and teaching purposes. There are also limited research on evaluation of generated questions by experts (teachers/educators) and on generating informative feedback \cite{kurdi2020systematic}.

\section{VidVersityQG - An AI-based solution for Digital Learning and Assessment} \label{sec:method}

\begin{figure}[t!]
   \includegraphics[width=1\linewidth]{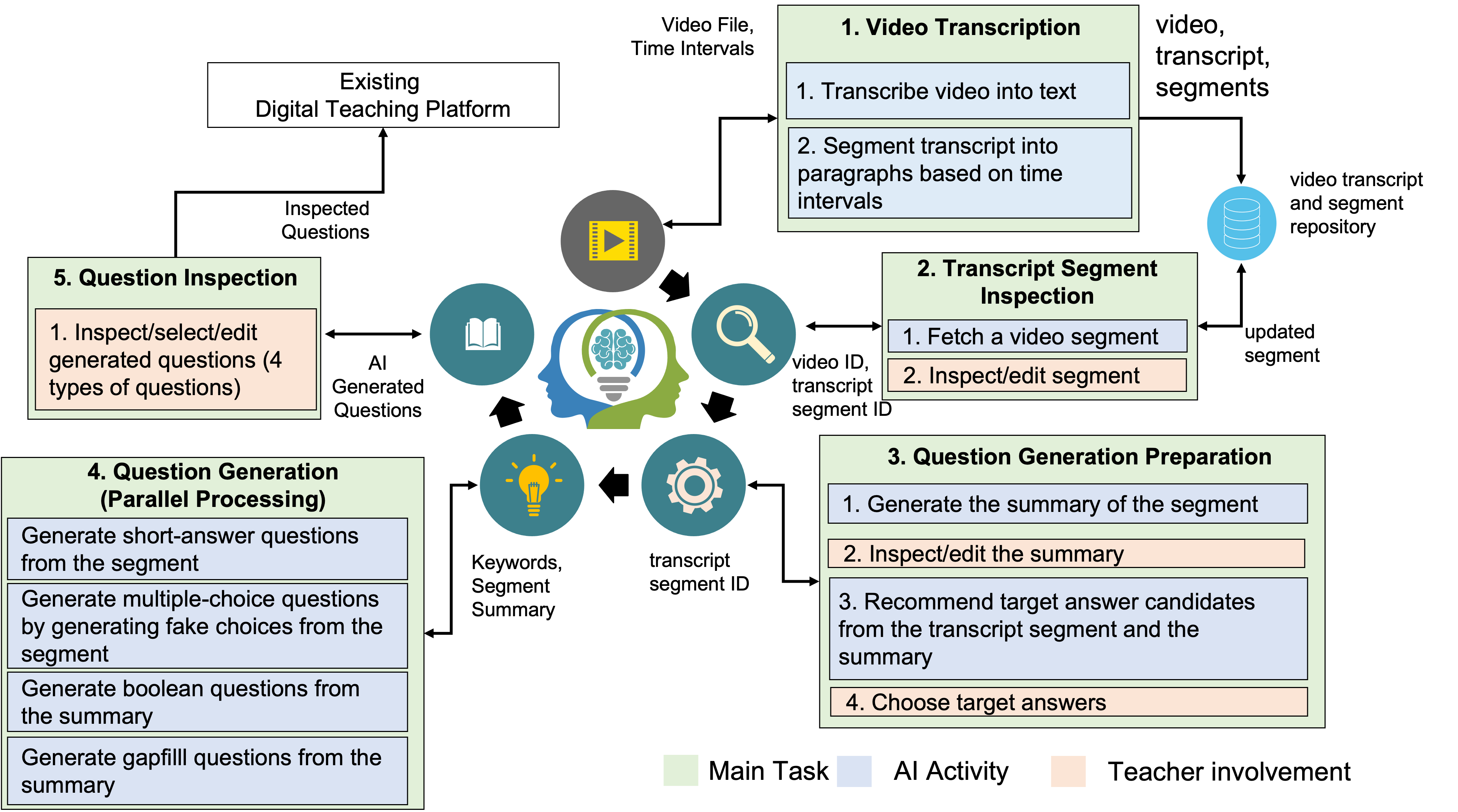}
    \caption{The pipeline of five phases in VidVersityQG}
  \label{fig:methodology}
\end{figure}

VidVersityQG is an AI-based solution that automatically generates natural language questions from a given educational video (e.g., lectures, guided instruction) to aid future teachers to continuously assess what learners understand. It follows a user-centered design approach enabling teachers to interact with the solution to refine the AI-generated questions. In VidVersityQG, digital learning and teaching involves five phases which have been jointly designed and developed with expert teachers/educators who bring significant experience in delivering digital learning and teaching outcomes. The five phases are presented in Figure \ref{fig:methodology} and described in the following subsections. 

\subsection{Video Transcription}

The VidVersityQG model of delivering digital learning and teaching is to divide pre-recorded video lectures into small conceivable chunks of videos (which we term as video segments). Such an approach for delivery of learning and teaching has been trialled and tested to be very effective by our industry partner VidVersity in both university and corporate settings. To support the above approach, in the first phase, there are two objectives. Firstly, automatically transcribe a given educational video to text (called \textit{transcript}) using an automatic speech recognition technique (e.g. Amazon Transcribe) \cite{awstranscribe}. Secondly, segment the transcript into smaller paragraphs based on the small conceivable video segments produced by the teacher/educator. Automatic video transcription significantly contributes to alleviating the time-consuming, expensive manual transcription task. 

In our proposed approach for learning and teaching, we aim to generate different questions from the transcript of each segment that can help to assess the different facets of the learned knowledge of a learner, based on context and content of the corresponding video segment. Generating questions from a paragraph is relatively difficult and considered a new research direction for automatic question generation \cite{das2021automatic}. This task requires a sufficient understanding of the larger context of the sentences included in each paragraph and the inner relation between the sentences using natural language understanding. 


\subsection{Transcript Segment Inspection}

The goal of the second phase is for teachers to inspect a “transcript segment” (simply referred to as segment) of a video to check for whether there are errors to be fixed. Our premise is that while a recent video transcription tool can be effective, it cannot replicate flawless speech-to-text conversions. In particular, since punctuation marks (e.g., comma, period, question marks) as well as pauses in speech when recording videos of training/lectures are usually not inferred accurately. Hence, the transcript generated in the first phase may contain some inaccurate usage of punctuation marks. To fill this gap as well as instil confidence and trust in the technology, teachers are involved in this task to make sure the transcript for the video segment is highly accurate. 


VidVersityQG has the ability to keep a track of every modification made by the teacher/educator to the transcript of the video segment, thus, if a mistake is made, teachers can retrieve earlier versions of the segment to help fix the mistake. These edits are also valuable source of data to further improve the AI algorithms ability to transcribe more accurately.

\subsection{Question Generation Pre-Processing}

In the third phase, we perform pre-processing to be able to generate question automatically in the next phase. The pre-processing is conducted with the involvement of the teacher/educator. 

The first pre-processing step involves generating a text summary of the transcript generated in the previous phase. 
This text summary will be used for generating both \textit{Gapfill} and \textit{Boolean} questions (Yes/No or True/False). 
Our key idea is to utilise an automatic text summarisation technique that aims to generate a smaller, easy-to-read and  concise piece of the original segment that conveys the key and relevant information of the segment \cite{hasan2021xl}. There are two main approaches to automatic text summarisation: \textit{extractive} and \textit{abstractive}. The extractive approach aims to generate a summary from an input text by extracting one or more segments from the text and aggregating them. This approach was popular in the earlier days of text summarisation, suffering from weak coherence between sentences and unitented repetition of sentences in the generated summary. The abstractive approach generates a summary that may contain words and phrases not present in the input text. It has been observed that these methods can produce better summaries that are closer to human-generated summaries than extractive methods \cite{hasan2021xl}. For this reason, we choose the latter approach. In particular, we use a recent pretrained language model, Google’s T5 (Text-To-Text Transfer Transformer) \cite{roberts2020exploring} (i.e., T5-base: one of the T5 models with 12 attention layers used to capture the meaning and position of each word in a given text) trained on an open-source dataset, the Colossal Clean Crawled Corpus (C4) \cite{c4dataset}. T5 has proven to be the state-of-the-art results on many NLP benchmarks in 2020 \cite{roberts2020exploring}.

Once a summary is generated from the chosen segment, teachers can be involved to check whether there is room for improving the quality of the summary. We note that while the current text summarisation models are good at generating abstractive summaries, they still may need to be improved by increasing language fluency and reducing grammatical errors \cite{roberts2020exploring}. The required human involvement is similar to what we have presented in the second phase. That is, teachers can edit and modify the summary, if necessary, and the changes made by the teacher are traceable and stored in VidVersityQG repository.

The second pre-processing task is essential for generating both \textit{short-answer} and \textit{multiple-choice} questions. 
To generate such question types, it is desirable that questions are relevant to the segment and target the key concepts (which we term as \textit{keywords}) that the teacher/educator wishes to assess the learner understanding. 
Thus, identifying the right keywords plays a significant role in question generation for these types of questions. That is, different questions can be generated from the same text based on the choice of keywords (which represent the concepts that need to be assessed). 
In our approach, a keyword can be either recommended by VidVersityQG's keyword generation module or chosen by the teacher. The keyword generation module automatically extracts a ranked list of both noun phrases and named entities from the segment. These phrases and entities are used as candidates for choosing keywords by a teacher. Noun phrases are defined as phrases that consist of a noun or pronoun and any number of constituents including adjectives, determiners, prepositional phrases, verb phrases, and adjective clauses. Named Entities are defined as the proper names identified in a text. Identified text may be a person‘s names, organisation‘s names, location‘s names, and date and time expressions. In the information retrieval discipline, nouns and named entities have been widely used to represent key concepts in text documents \cite{kang2014cfinder}. Our solution presents a ranked list of these candidates, based on their frequencies in the segment, to a teacher. Then, the teacher can choose multiple ones from the list as keyword. Alternatively, a teacher can manually choose a set of words from the segment transcript as the target keywords. 

The final pre-processing step is to choose a set of keywords from the summary generated in the first step of pre-processing. Choosing these keywords is a prerequisite for generating \textit{gapfill} questions. Similar to the previous pre-processing step, VidVersityQG can automatically generate some keywords that can be used to assess the learners understanding of the concept delivered in the video segment or allow the teacher/educator to make a selection.

\subsection{Question Generation}

In this section, we present the proposed AI-based NLP model for generating four types of questions: short-answer, multiple-choice, boolean and gapfill question types.

\subsubsection{Short-answer question generation}

In a short-answer question, we expect a learner to provide a word or phrase as a response to a question. Ideally, the learner-provided answer must match or needs to be highly similar to the target answer (i.e., ground-truth answer) chosen by a teacher. VidVersityQG generates short-answer questions from the two input kinds: (1) the target segment chosen by a teacher in the second phase, and (2) the keywords chosen in the third phase. 
We employ an advanced NLP model that uses a deep learning neural network, named ParaQG \cite{kumar2019paraqg}, to automatically generate short-answer questions. ParaQG can generate fluent, meaningful and relevant questions from a paragraph with a given target answers. From the given segment and each keyword, we generate a ranked list of questions, where the ranking is determined by confidence scores produced by the model. By default, the top-3 questions are presented to the teacher, which the teacher can make edits if necessary. More details about the process are described in the Question Inspection section below.

\subsubsection{Boolean question generation}

Assessing whether learners understand what key factual statements are true or false from education materials (in our case, transcripts) is an essential part in promoting the quality of education. To achieve this, our aim is to automatically generate boolean questions that require a learner to provide Yes/No or True/False answers. In the context of this study, desirable boolean questions need to go well beyond what is immediately stated in a given transcript and assess the overall comprehension of the learner about key information delivered from the transcript. To automatically generate boolean questions, an important challenge is how a machine can {infer} high-quality Yes/No questions using AI techniques from a transcript. 

To tackle this challenge, our solution is equipped with a boolean question generation module that uses state-of-the-art inference abilities. Our key idea is to generate boolean questions from the \textit{summary} of a given transcript. Recall that our summary generation is based on abstractive text summarisation that can generate a short and concise summary that captures key ideas of a transcript, as presented in phase 3. Given the summary, we use the T5-base model \cite{roberts2020exploring} trained on a recent boolean question generation dataset, named BoolQ \cite{clark2019boolq}, which contains nearly 16k Yes/No question-answer examples. The T5 models were designed to be used for a variety of NLP tasks such as text translation, summarisation, and question generation and have shown outstanding performance for high-quality Boolean question generation \cite{roberts2020exploring}. Using the T5-base model, we generate two sets of top-\textit{N} boolean questions (\textit{N}: a teacher-specified parameter), where one set requires Yes (or True) and the other set requires No (or False) as the correct answer. 

\subsubsection{Gapfill question generation}

Gapfill (also called ``fill-in-the-blank'') questions are also popularly used to assess learners' knowledge, where learners are asked to fill one or more omitted words given a text. In our context, our strategy is to automatically generate gapfill questions from the summary of the chosen transcript. The reason is that if we generate this type of questions from a transcript, the majority of the sentences of the transcript may not be appropriate for generating high-quality questions \cite{das2019automatic}. Since the corresponding summary contains key information of the transcript, we choose to use it as the primary information source for generating gapfill questions. The omitted words are chosen in the ``question generation preparation'' phase by a teacher, as mentioned earlier, and they will be used as the correct answers. To sum up, the summary itself with omitted words is presented to a learner as a gapfill question, requiring them to provide the omitted blanks (or gaps) to complete the summary. 

\subsubsection{Multiple-choice question (MCQ) generation}

MCQs are a popular form of assessment in which the learners are asked to choose the best possible answer(s) out of a few choices. A set of choices is the fundamental requirement for generating a MCQ. The actual answers to the question must be included in those choices. The incorrect choices are often called distractors. 
In our approach, MCQs are automatically generated for assessing specific knowledge embedded in a small sized text (e.g., a sentence). Moreover, MCQs can be generated from key, factual information conveyed through the transcript. To achieve the former objective, we use the short-answer question generation module with our distractor generation module. For the latter case, we use the gapfill question generation module with the distractor generation module. In both cases, a common, key challenge is how to generate good candidates of distractors. To address this, we use the recent distractor generation model \cite{chung2020bert} that uses a state-of-the-art neural network. The goal of this model is to generate multiple context-related incorrect choices. This model was trained on a large-scale question-answer dataset, called RACE \cite{lai2017race}, that contains more than 27k text articles with more than 97l questions from English examinations of middle-school and high-school Chinese students within the 12-18 age range. Given the input of a transcript and the correct answer, this model can generate top-\textit{N} (\textit{N}: a user-specified parameter) distractors. 

\subsection{Question Inspection}

The final phase of VidVersityQG is to inspect four types of questions generated in the previous phase by teachers. Note that generated questions are ranked in terms of their confidence scores determined by the question generation modules in VidVersityQG. Teachers can examine the questions closely to assess their quality. If necessary, they can manually modify to make the questions more suitable for use by learners. Finally, teachers can select the questions to be used and save them to the VidVersityQG repository which can then be presented to the learner via the existing digital learning and teaching delivery platform (see Figure \ref{fig:methodology}).

\section{Results}

In this section, our aims are two-fold. The first is to demonstrate how VidVersityQG can facilitate technology-enhanced learning. The second aim is to report the human-based evaluation results of VidVersityQG in terms of its effectiveness. To achieve the first aim, we show the functionalities of VidVersityQG based on its Web-based user interface (UI). The second aim is accomplished by evaluating the quality of generated questions by 7 experienced teachers using the teaching/training videos obtained from 12 domains including aged care, language education (English), law, banking and finance, leadership, business development, safety training, well being, workplace education, hope science, marketing and human factors.

\subsection{VidVersityQG - Web UI Demonstration}

We have co-designed a Web-based UI in consultation with teaching professionals working with VidVersity \cite{vidversity2021}. These professionals have extensive years of experience in learning and teaching in both university and corporate environments. The UI aims to enhance the efficiency of teachers and it provides an easy-to-follow step-by-step guide for question generation by VidVersityQG core modules. In addition, as discussed in the previous  section, it offers the teacher/educator the  option to verify the correctness and quality of generated questions and to make necessary modifications if required. Further, at any phase of the question generation process, teachers  can go back and make necessary adjustments.

\subsubsection{Video Transcription}

To generate questions, the pre-requirement is to transcribe learning/training videos. In the VidVersityQG implementation, we used AWS Transcribe \cite{awstranscribe}  for this purpose. Given a video, it extracts speech information and converts it into text using deep learning techniques. Then, the text is segmented into smaller transcripts, where each transcript is used as an input for question generation in VidVersityQG. For segmentation, VidVersityQG can incorporate two options. The first is to segment the text based on a fixed time duration (e.g., 5 mins), and another option is to segment it based on the input from teachers. The video transcription phase is not part of the UI, and is performed in a batch mode with an python Application Programming Interface (API). Multiple videos can be transcribed together in VidVersityQG.

\subsubsection{Transcript Segment Inspection}

\begin{figure}[t!]
    \includegraphics[width=0.95\linewidth]{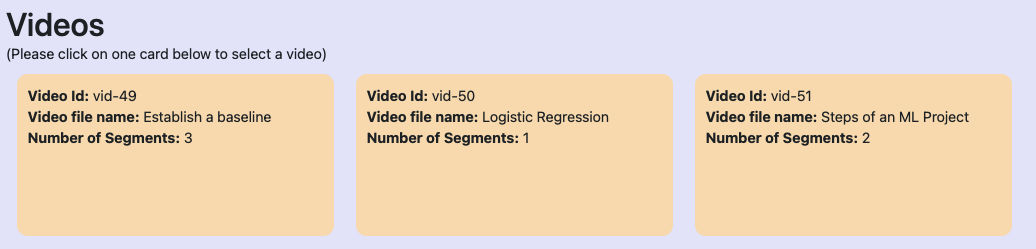}
    \caption{Video selection for question generation}
    \label{fig:videos}
\end{figure}

\begin{figure}[t!]
    \includegraphics[width=0.65\linewidth]{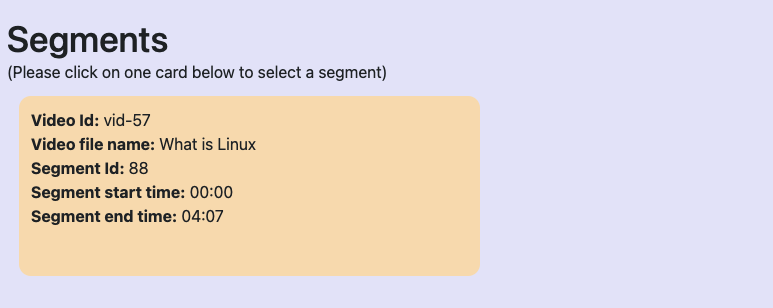}
    \caption{Segment selection for question generation}
    \label{fig:segment}
\end{figure}

\begin{figure}[t!]
    \includegraphics[width=0.95\linewidth]{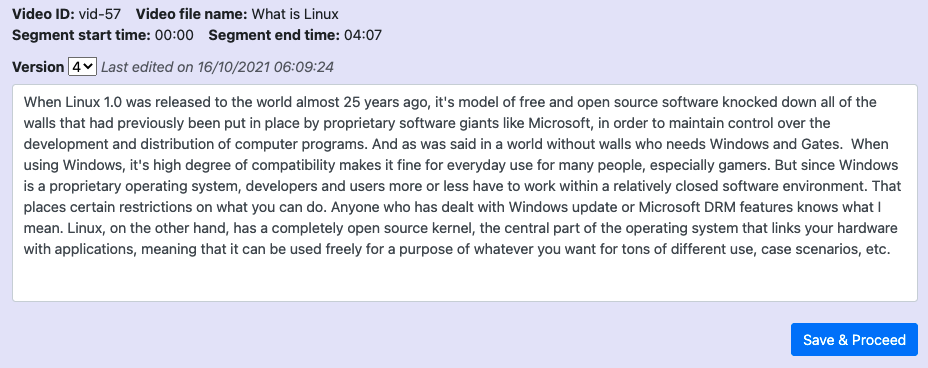}
    \caption{Transcript segment inspection and tracking version trajectories}
    \label{fig:inspect_transcript}
\end{figure}

Figure \ref{fig:videos} shows a snapshot of the UI that shows a list of the transcribed videos with their titles and corresponding number of segments. Once a video is selected, the teacher can choose a segment to be used for question generation as seen in Figure \ref{fig:segment}. Then, the teacher can inspect the transcript for the corresponding video segment using the UI as depicted in Figure \ref{fig:inspect_transcript}. As seen, the teacher can review the segment’s text and make necessary edits (e.g., correct grammar, punctuations, missing words in the transcribed text, etc.), and then save this. Once an updated segment is saved, the teacher can access and reload any previously saved version of the segment by choosing one from the “Version” dropdown list box. Also, the UI shows the last edited date and time information of each version to help the teacher to track the trajectory of the past versions.

\subsubsection{Question Generation Preparation}

\begin{figure}[t!]
    \includegraphics[width=0.80\linewidth]{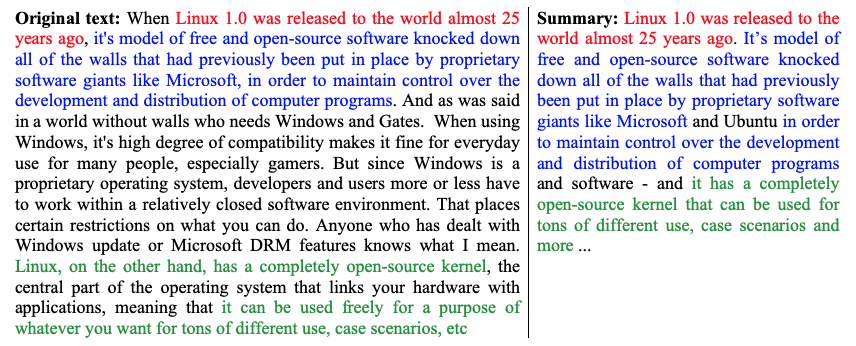}
    \caption{Segment summary generation}
    \label{fig:ori_summary}
\end{figure}

\begin{figure}[t!]
    \includegraphics[width=0.95\linewidth]{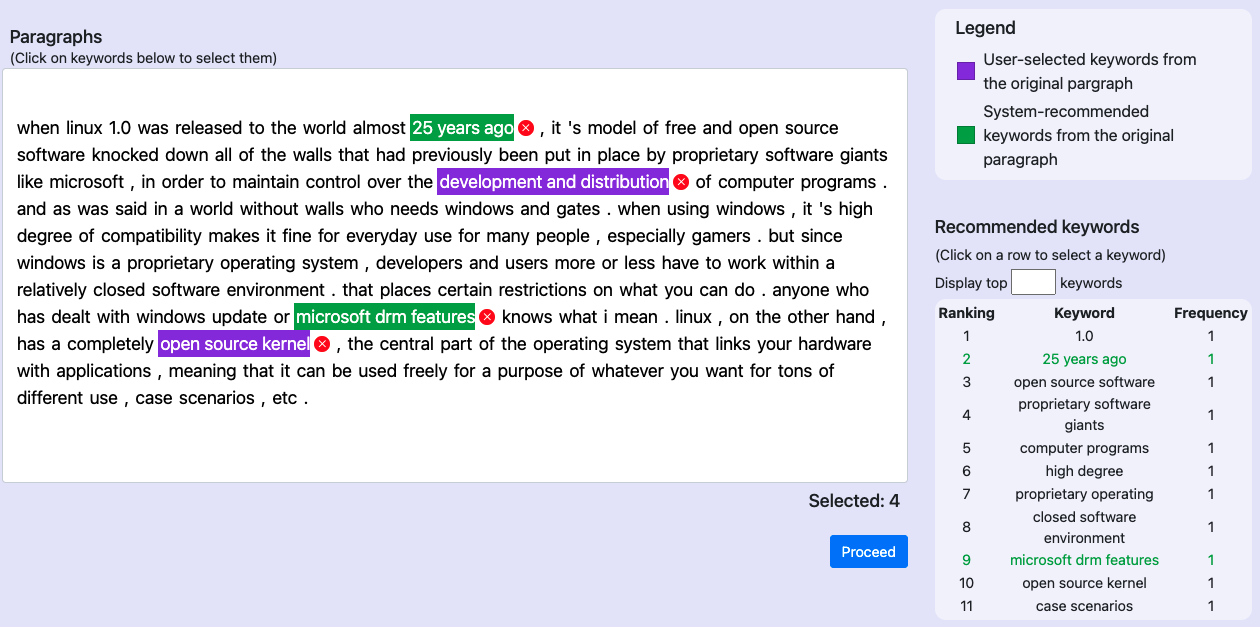}
    \caption{A ranked list of recommended keywords and chosen ones by a teacher (green color). The custom keywords can be also chosen by the teacher based on the teacher’s expertise purple color).}
    \label{fig:saq_mcq_kw}
\end{figure}

Figure \ref{fig:ori_summary} shows an example of a summary generated from a segment. The teacher can inspect the summary, and make any modifications if necessary. Modification made to summary are tracked and versions stored in the VidVersityQG repository.
Figure \ref{fig:saq_mcq_kw} (right) shows an example of a list of ranked recommended keywords based on their frequencies generated by the keyword generation module. As seen, the teacher can refer to this list and choose either the recommended keyword by VidVersityQG or can manually select the keywords. The recommended keywords are presented in green colour while the user selected one are highlighted in purple colour. 
As the final step in this phase, the teacher can see the text summary of the segment and choose keywords for generating gapfill questions using a similar interface with Figure \ref{fig:saq_mcq_kw}.


\subsubsection{Question Generation}

\begin{figure}[htt!]
    \includegraphics[width=0.95\linewidth]{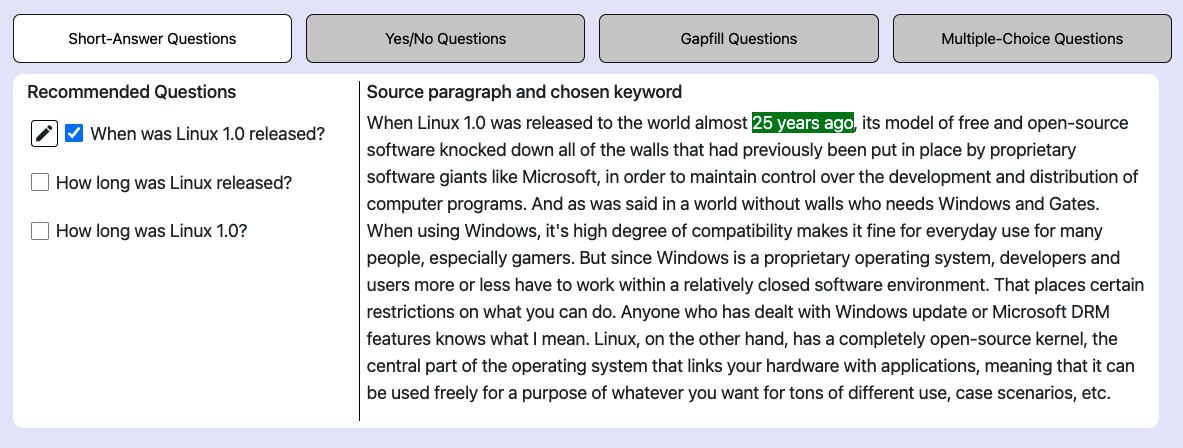}
    \caption{An example of short-answer question (SAQ) generation}
    \label{fig:saq}
\end{figure}

\begin{figure}[ht!]
    \includegraphics[width=0.95\linewidth]{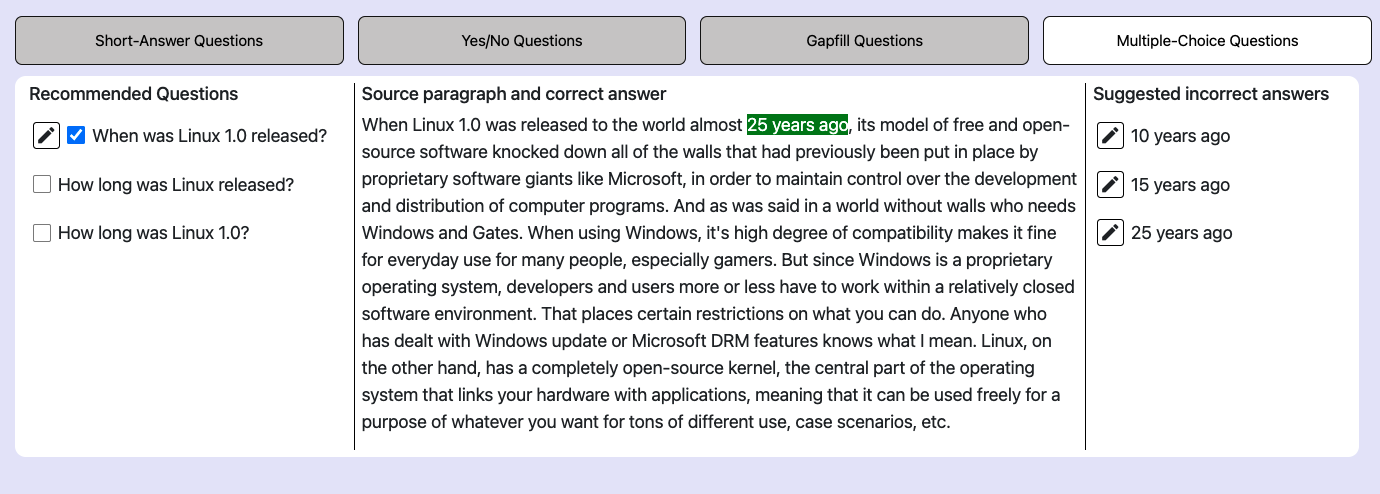}
    \caption{An example of multiple-choice question (MCQ) generation}
    \label{fig:mcq}
\end{figure}

VidVersityQG can generate four different types of questions: short-answer questionss (SAQs), boolean questions (BLQs), gapfill questions (GFQs) and multiple-choice questions (MCQs). Figure \ref{fig:saq} shows three generated SAQs (in the left panel) for one keyword selected in the question preparation stage (see also Figure \ref{fig:saq_mcq_kw}) that is highlighted in green colour. Similarly the teacher can see the same information for MCQs with the three suggested incorrect answers such as ``10 years ago'', ``15 years ago'' and ``20 years ago'' (see Figure \ref{fig:mcq}).

\begin{figure}[ht!]
    \includegraphics[width=0.95\linewidth]{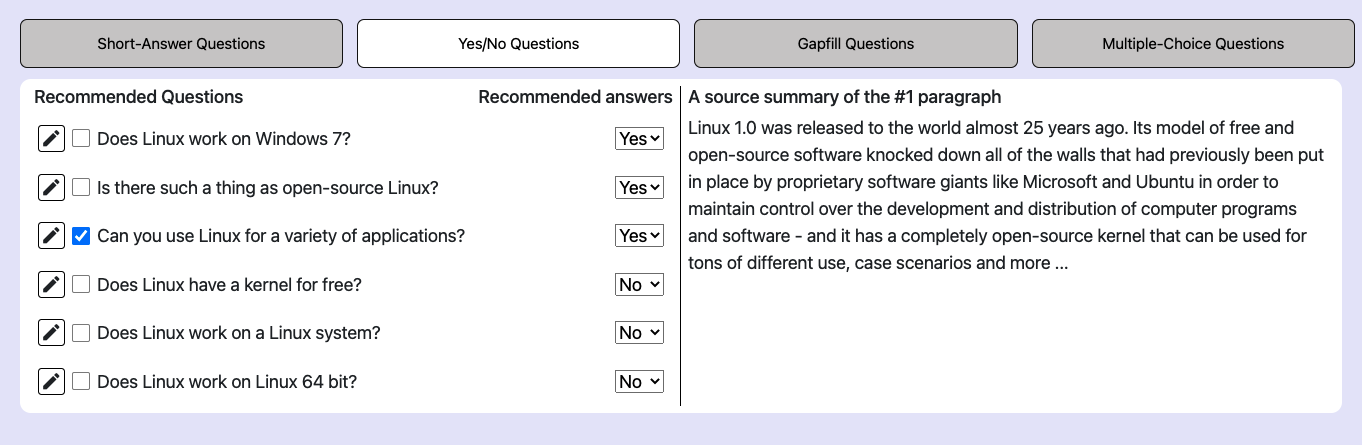}
    \caption{An example of boolean question (BLQ) generation}
    \label{fig:boolq}
\end{figure}

\begin{figure}[ht!]
    \includegraphics[width=0.95\linewidth]{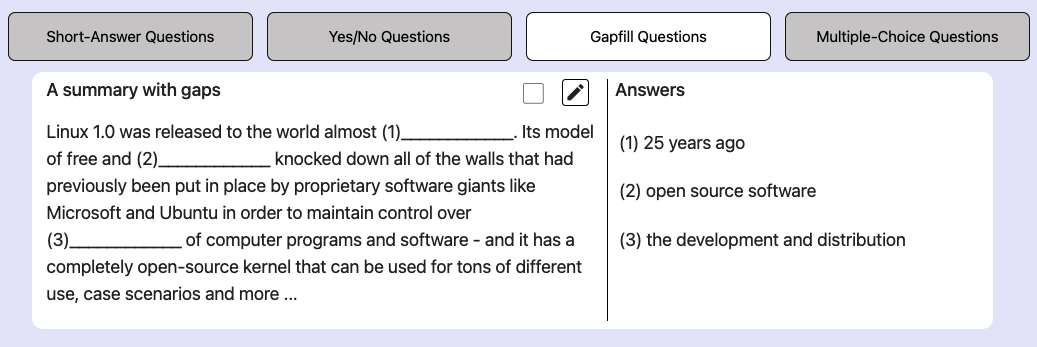}
    \caption{An example of gapfill question (GFQ) generation}
    \label{fig:gf}
\end{figure}

For BLQs and GFQs, questions are generated via the summary text as shown in Figure \ref{fig:ori_summary}. Regarding BLQs, we present three questions whose answer is ``Yes'', three questions whose answer is ``No'', and the summary text used to generate the questions (shown in Figure \ref{fig:boolq}). In terms of GFQs, we present the summary text with gaps and numbers that indicate the order of the answer for each gap. As shown in Figure \ref{fig:gf}, there are three gaps corresponding to three answers such as ``25 years ago'', ``open source software'' and ``the development and distribution''. 

\subsubsection{Question Inspections and Modifications}

\begin{figure}[ht!]
    \includegraphics[width=0.95\linewidth]{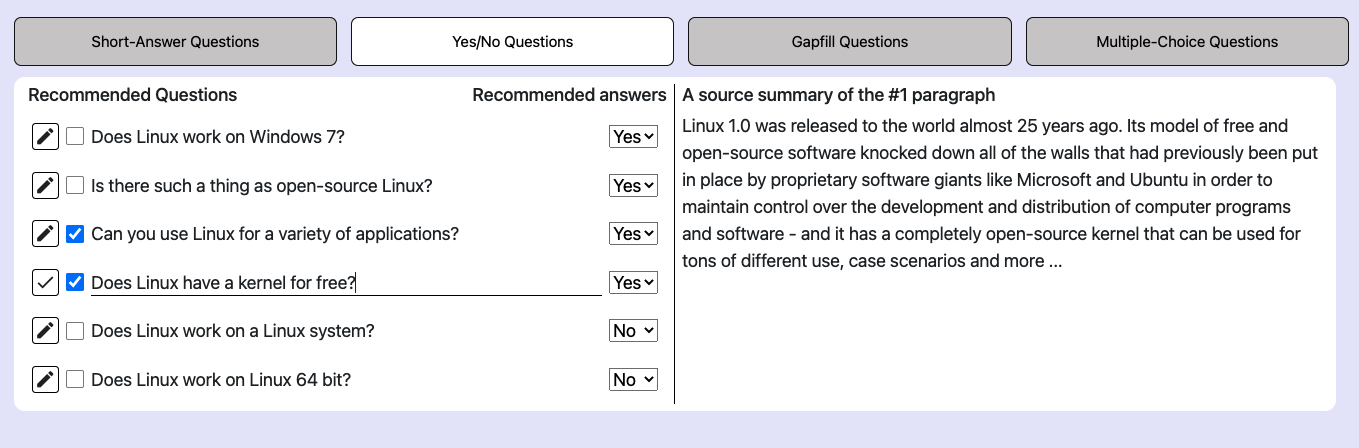}
    \caption{An example of BLQ inspection}
    \label{fig:inspect_questions}
\end{figure}

There are four buttons located in the top to switch between four question types (as shown in Figure \ref{fig:saq}). The teacher can select a question type, get option to edit and save it. Figure \ref{fig:inspect_questions} shows how to inspect and edit boolean questions. Here, the teacher can see the generated questions on the left panel that can be edited by clicking the edit icon and the summary text from the source paragraph in the right panel. Furthermore, the teacher can change the recommended answer using the dropdown that contains either the ``Yes'' option or the ``No'' option. Similarly, the teacher can inspect SAQS (Figure \ref{fig:saq}), GFQS (Figure \ref{fig:gf}) and MCQs (Figure \ref{fig:mcq}).

\subsection{VidVersityQG - Evaluation Setting}

We evaluated VidVersityQG using real-world online educational learning/training videos. These videos were chosen from 12 different domains  such as law, business management, aged care, workplace training, English language and others
to comprehensively evaluate the quality of generated questions by VidVeristyQG. 

\subsubsection{Data Preparation}
Assessing  the quality of generated questions in each domain was performed by 7 experienced teaching professionals, where 3 professionals are from business domains, 2 professionals are from education domains, and the remaining 3 are from law domains. In total, 117 education videos were used. The shortest video length is 24 seconds and the longest video length is 57 minutes and 13 seconds. The average length of a video is around 12 minutes. For each video, we applied the segmentation process to divide them into  smaller video segments where each video segment has a maximum length of 5 minutes. For example, a video with a length of 7 minutes is sequentially cut into two segments such as the 5-mins and 2-mins segments. In total, we have 247 video segments generated from those 117 videos with a video segment length from 52 seconds to 5 minutes. The average length of all segments is 4 minutes and 35 seconds. All 7 professionals were instructed to select and thoroughly watch the videos relevant to his/her expertise before generating questions by VidVersityQG using the UI. Also, they were asked to look at generated segments of chosen videos, generate questions using VidVersityQG and evaluate their quality, if the segments were worth for generating questions. Because we noticed that all segments may not have important educational information that can be used to assess the learner's understanding and proficiency. Finally, 52 segments of 46 videos were chosen for evaluation purpose by the 7 professionals. We discuss the evaluation outcome for the generated question from these segments.  

\subsubsection{Evaluations settings}
To assess the  quality of the recommended top-3 SAQs by VidVersityQG for a keyword, each professional were asked to give a rating from the following three categories:

\begin{itemize}
    \item \textbf{Good}: If at least one of the recommended questions is acceptable without modification or with minor modifications
    \item \textbf{Average}: If at least one of the recommended questions is acceptable with reasonable modifications
    \item \textbf{Bad}: If none of the recommended questions are acceptable and require major modifications
\end{itemize}

\begin{figure}[ht!]
    \includegraphics[width=0.95\linewidth]{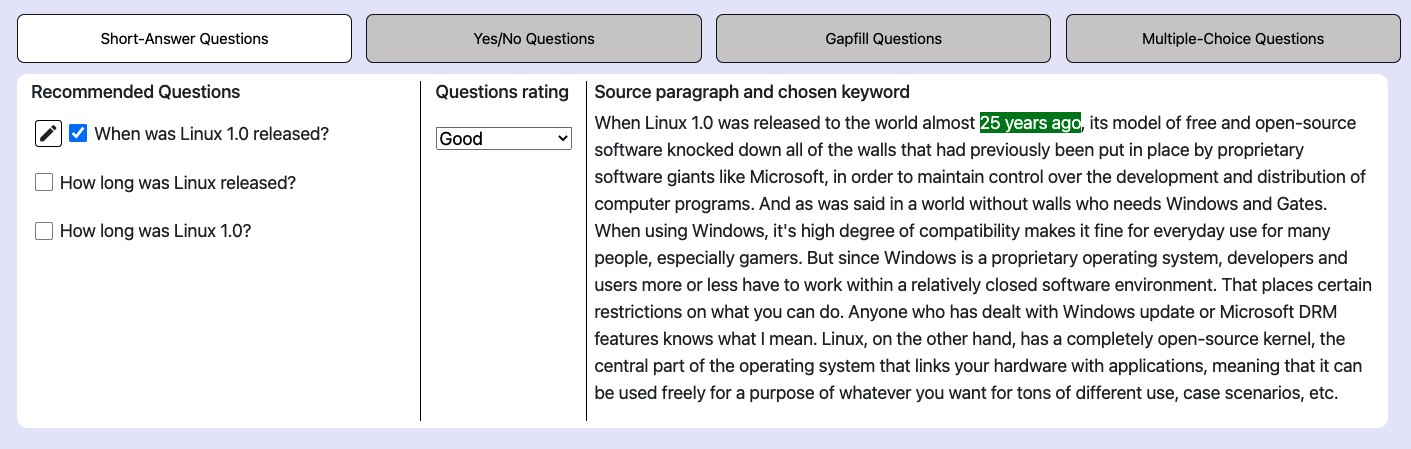}
    \caption{Assessment of the top-3 SQAs using the target keyword in green colour from a segment.}
    \label{fig:saq_eval}
\end{figure}

Figure \ref{fig:saq_eval} is an example that shows the assessment of the top-3 SAQs using the target keyword highlighted in green colour from the segment. The three categories of a rating option were presented in a dropdown list box. If a ``Good'' rating is given, the professional was also asked to select the best question. In this example, the first question was chosen as the best question.

\begin{figure}[ht!]
    \includegraphics[width=0.95\linewidth]{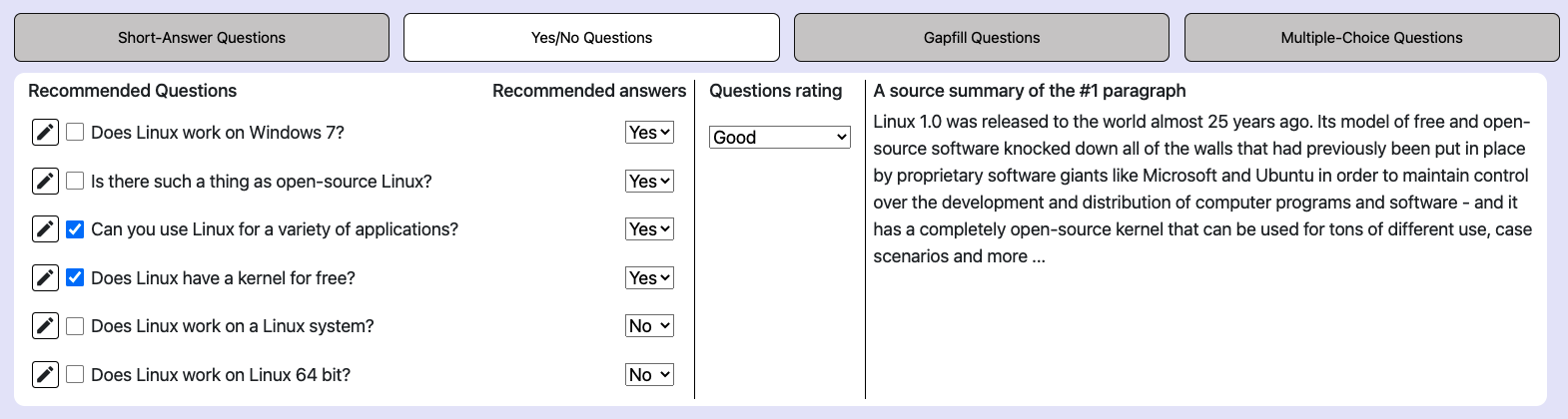}
    \caption{Assessment of the six BLQs from the summary of a segment}
    \label{fig:boolq_eval}
\end{figure}

To assess BLQs, the top-3 BLQs (whose ground truth answer is ''yes'') and the other top-3 BLQs (whose ground truth answer is ``no'') were given to each professional for evaluation from the summary of the chosen segment. To assess the quality of the six automatically generated questions, each professional was asked to give a rating, ``Good'', ``Average'' and ``Bad'' based on the same criteria used for assessing SAQs. Figure \ref{fig:boolq_eval} shows an example of the UI used for assessment.

Each professional assessed MCQs by giving a rating based on the quality of distractors related to each MCQ. To achieve this, we used the same SAQs as MCQs, thus the only difference between SAQs and MCQs is that MCQs are the same SAQs with three distractors. Therefore, for each keyword chosen by each the professional, we used the top-3 SAQs, and the professional was asked to give a rating based on these criteria:

\begin{itemize}
    \item \textbf{Good}:  If the best question’s distractors are acceptable with minor modifications
    \item \textbf{Average}: If the best question’s distractors are acceptable with reasonable modifications
    \item \textbf{Bad}: If the best question’s distractors are non-acceptable. 
\end{itemize}

\begin{figure}[ht!]
    \includegraphics[width=0.95\linewidth]{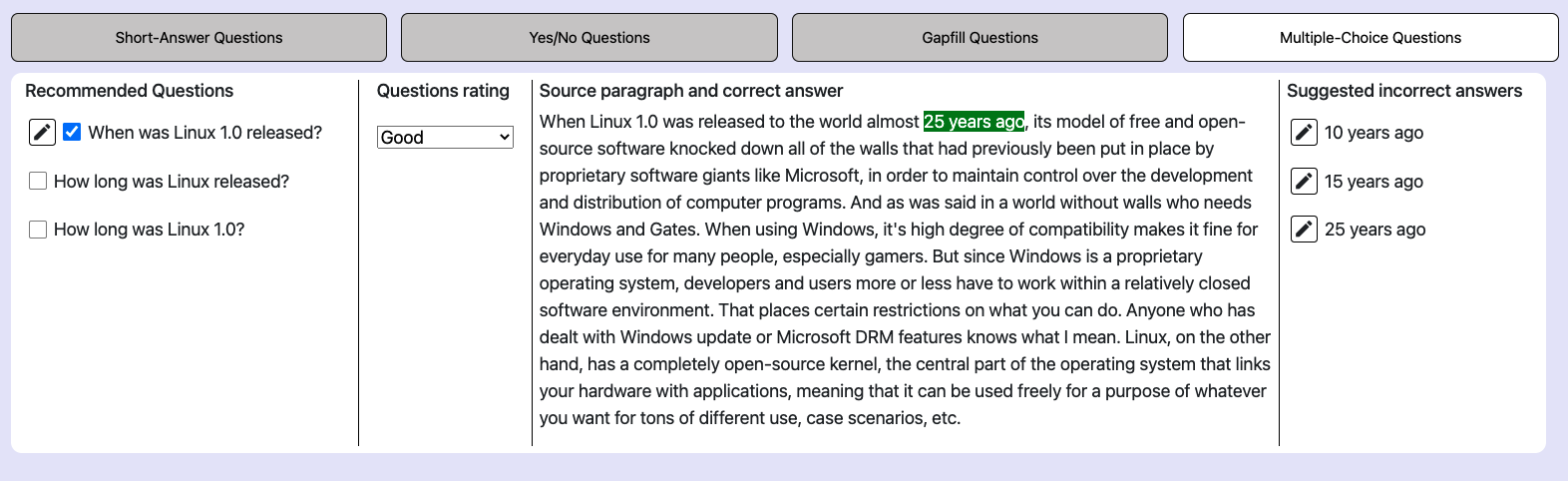}
    \caption{Assessment of the MCQ using three generated distractors}
    \label{fig:mcq_eval}
\end{figure}

An example of this assessment is shown in Figure \ref{fig:mcq_eval}. This shows that the first question was chosen as the best question with the three distractors (``10 years ago", ``15 years ago", and ``20 years ago"), where the target keyword was chosen as ``25 years ago” (as highlighted in green).

Finally, to assess GFQs, we asked each professional to assess the quality of the summary of the segment. The reason is that given a summary, a GFQ is generated by omitting target keywords from the summary, where such keywords were chosen by the professional. In other words, the summary itself with omitted words is presented as a GFQ. The professional was asked to give a rating based on these criteria:

\begin{itemize}
    \item \textbf{Good}: If the summary is acceptable with  minor modifications
    \item \textbf{Average}: If the summary is acceptable with reasonable modifications
    \item \textbf{Bad}: If the summary is non-acceptable. 
\end{itemize}

\subsection{VidVersityQG - Evaluation Outcome}
\begin{figure}[t!]
    \includegraphics[width=0.95\linewidth]{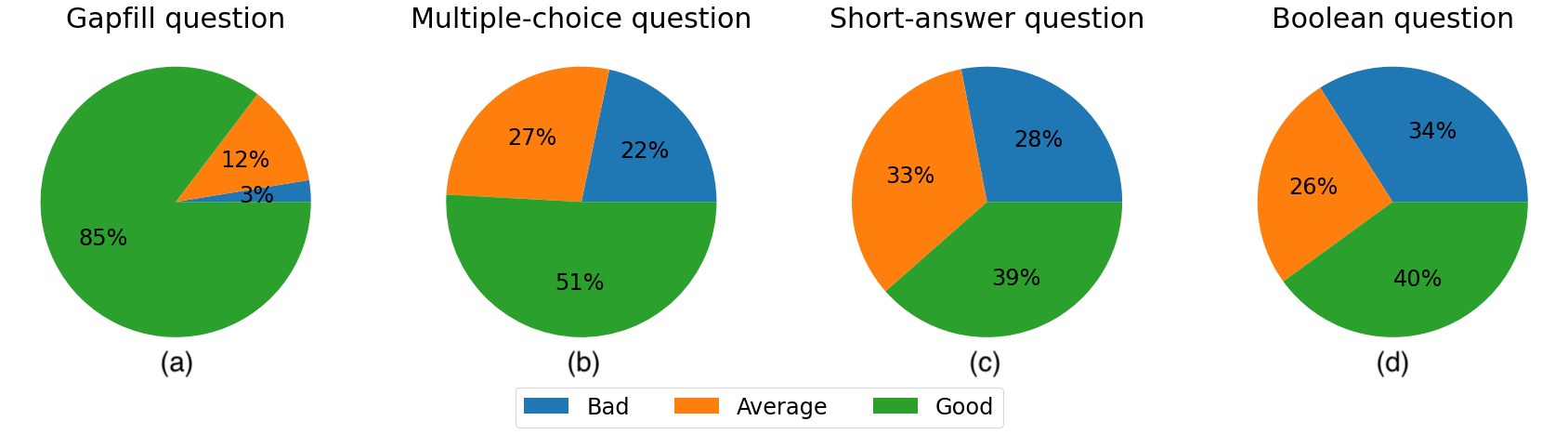}
    \caption{The quality assessment of generated questions of four types.}
    \label{fig:question_quality}
\end{figure}

The total of the 335 SAQs, 116 GFQs, 346 MCQs and 164 BLQs were generated and evaluated by the 7 professionals. Figure \ref{fig:question_quality} presents the assessment results that show the proportion of the given ratings by the professionals on the four types of questions. The following key observations from this evaluation are drawn:

\begin{itemize}
    \item The proportions of the questions rated with ``Good'' are dominated in all four types of questions: 85\% in GFQs, 51\% in MCQs, 39\% in SAQs and 40\% in BLQs.
    \item The acceptable questions (rated with ``Good'' and ``Average'') are proven to cover 96\%, 78\%, 72\% and 66\% proportions in GFQs, MCQs, SAQs and BLQs, respectively.
\end{itemize}

\begin{figure}[ht!]
    \includegraphics[width=0.95\linewidth]{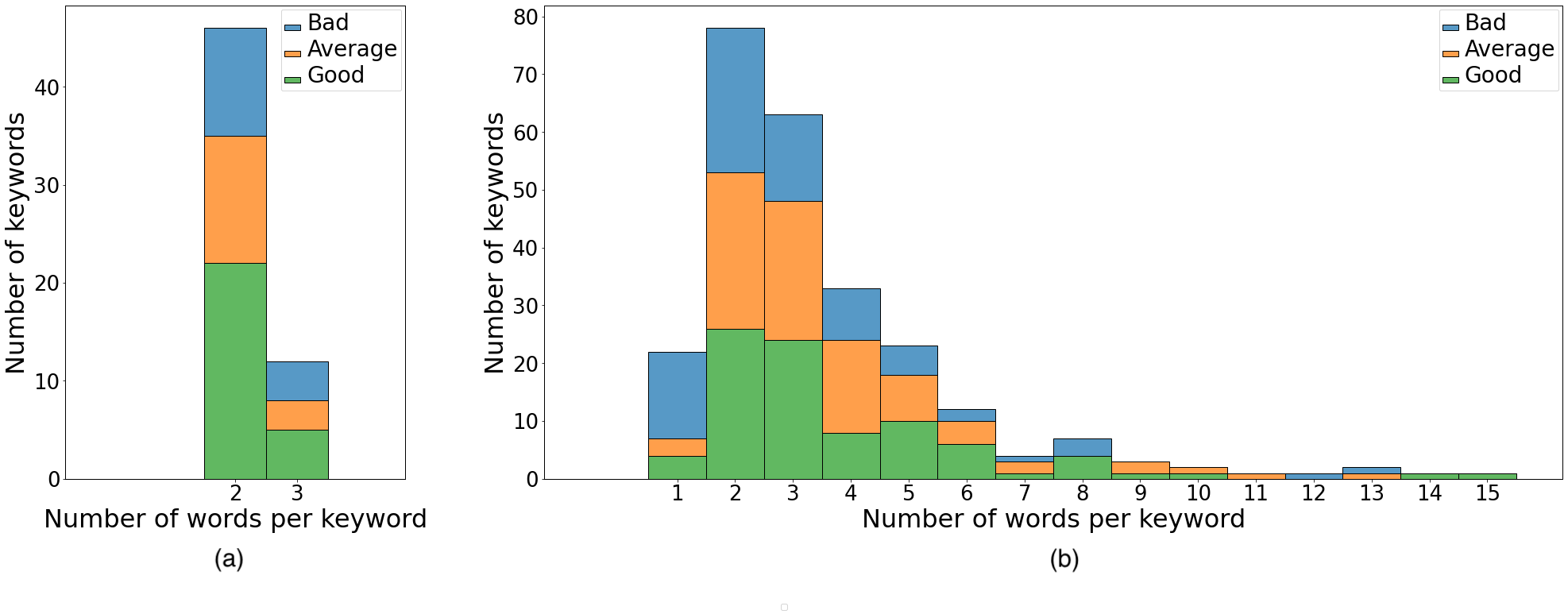}
    \caption{The distributions of the number of words per keyword used to generate SAQs: (a) keywords recommended by VidVersityQG, (b) custom keywords selected by the 7 professionals.}
    \label{fig:question_kw}
\end{figure}

SAQs often provide the opportunity to better assess the  learner's understanding and proficiency about the learning materials. Thus, we analyse the assessment results on SAQs more deeply. In particular, we are interested in understanding and analysing the relationship between the quality of generated questions and the number of words in target keywords used to generate the SAQs. The analysis outcomes are  presented in Figure \ref{fig:question_kw}. We observer the following points from the results:

\begin{itemize}
    \item As seen in Figure \ref{fig:question_kw}(a), the professionals mostly preferred to use 2 word-length of recommended keywords for SAQ  generation. On the other hand, it turned out that they dominantly used 2 and 3 word-length of custom keywords that were chosen by the professionals as we see in Figure \ref{fig:question_kw}(b).
    \item The professionals used custom keywords more than recommended keywords to generate SAQs. However, recommended keywords indeed can help them to identify potential target answers.
    \item Intuitively, the ``Good'' ratings and acceptable ratings (``Good'' and ``Average'') are not correlated to the word-length of keywords, as their proportions do not significantly change over different word-lengths on the x-axis.
\end{itemize}

\section{Discussion}

For years it has been predicted that AI will play a key role in assessment in education. However, much of the literature concludes the impact has been limited and not reached expectations. One of the key reasons has been the absence of a human centred approach where the AI assists the educators to create questions and assess the students answers. It has lacked the ability to adapt to the human requirements. This paper demonstrates how AI, specifically NLP when co-designed with teachers/educators can be used to effectively deliver digital learning and teaching while aiding teachers in assessing and understanding the knowledge of the learners. Our proposed VidVersityQG digital learning and teaching approach allows teachers/educators to provide for continuous assessment at scale, opening the door to AI reaching its potential as a tool for future teachers to enhance delivery of digital learning.  

Our approach has been designed to be generalisable, aiming to apply to education domains as well as industry sectors where video-based training is essential, e.g.,  training for safety of workers and training for improving skills of employees. We believe that the proposed method can provide two major benefits for future education community: The first is to give educators the ability to quickly and easily  generate  assessment tasks for learners and to scale this process to large cohort of learners. One of the most difficult aspects of scaling online education is the assessment component. An educator can record a lecture or create content that can be delivered to thousands of learners (the one to many approach) but assessing learners is extremely time consuming. The method discussed here provides a ‘digital assistant’ to the educator that helps generate questions automatically to assesses the learner. This is can be done quickly by taking the transcript from a video, creating questions using keywords generated by the AI and/or the educator and provide instant feedback to the learner. The second benefit (and part of our future work) is to enable personalised learning streams for the learner by adapting assessment question styles and content  based on their individual learning needs. Personalised learning pathways are known to be extremely beneficial to learner and allows them to learn at a pace suitable to their own needs.  By using different keywords, learner can be assessed with customised questions without having to recreate separate content. This method will enable educators to easily create personalised learning paths at scale.

There may be some possible limitations of this paper. In order to generate questions, this paper has proposed the use of the state-of-the-art deep learning algorithms. The effectiveness of such algorithms partially depends on the quality and amount of the training data used to build them. Most of the training data, currently available to use, are still limited and not categorised by specific education sectors. Thus, it would be promising to continue to retrain (or rebuild) the algorithms on more comprehensive data, if available in the future, to improve their effectiveness. In addition, to validate the quality of our approach more accurately, we may need to undertake more rigid and large-scale evaluation. Also, more case studies may need to be performed in more education domains to validate the practicability of the proposed approach by including various learners and teachers in the future. 

\section{Conclusion}
This paper presented a promising AI-based solution for advancing digital learning on the basis of video education materials by semi-automatically generating different types of questions using both AI and educators (or teachers). 
We highlighted the design, motivation, development and assessment of the proposed solution in detail. The key feature of our AI-based solution is to incorporate the state-of-the-art deep learning  and natural language process techniques into different types of flexible question generation with educator knowledge. We showed a great potential of our solution not just for improving time efficiency of conventional manual-based question generation, but also for harnessing current AI techniques to generate high quality questions from segment-level text snippets automatically captured from video teaching materials.
In the past few decades, we have seen a variety of question generation algorithms in computer science, however, their practicability and effectiveness have been little explored in education domains. This paper presented how to fill this gap with the proposed AI-based solutions for enhancing delivery of learning for future teachers.

\section*{Abbreviations}
\hfill \break
\begin{tabular} { | p{2.5cm} | p{8cm} | }
    \hline
    Abbreviation & Meaning \\
    \hline\hline
    AI & Artificial Intelligence \\
    \hline
    API & Application Programming Interface \\
    \hline
    AQG & Automatic Question Generation \\
    \hline
    AWS & Amazon Web Services \\
    \hline
    BERT & Bidirectional Encoder Representations from Transformers \\
    \hline
    BLQ & Boolean Question \\
    \hline
    C4 & Colossal Clean Crawled Corpus \\
    \hline
    DL & Deep Learning \\
    \hline
    DNN & Deep Neural Network \\
    \hline
    E-Learning & Electronic Learning \\
    \hline
    Gapfill & Fill in the blank question \\
    \hline
    GFQ & Gapfill Question \\
    \hline
    GPT-2 & Generative Pre-trained Transformer 2 \\
    \hline
    GPT-3 & Generative Pre-trained Transformer 3 \\
    \hline
    ITS & Intelligent Tutoring Systems \\
    \hline
    MCQ & Multiple-choice Question \\
    \hline
    MIT & Massachusetts Institute of Technology \\
    \hline
    MOOCs & Massive Open Online Courses \\
    \hline
    NLP & Neuro-linguistic programming \\
    \hline
    SAQ & Short Answer Questions \\
    \hline
    Seq2Seq & Sequence-to-sequence \\
    \hline
    T5 & Text-To-Text Transfer Transformer \\
    \hline
    TAM & Teacher Acceptance Model \\
    \hline
    UI & User Interface \\
    \hline
\end{tabular}

\section*{Acknowledgements}
This work was supported by a grant provided by VidVersity. We especially thank Simon Quirk (CEO of VidVersity) and Mike Levi (Technology Lead at VidVersity) for providing insightful comments and support during the project.

 \bibliographystyle{elsarticle-num} 
 \bibliography{main}





\end{document}